\documentclass[twocolumn,showpacs,amsmath,amssymb,superscriptaddress]{revtex4}
\usepackage{epsfig}
\usepackage{amsmath}
\usepackage{amssymb}
\usepackage{graphics}
\usepackage{xcolor,soul}
\usepackage[colorlinks=true,citecolor=blue,linkcolor=blue]{hyperref}

\DeclareMathAlphabet{\bi}{OML}{cmm}{b}{it}

\begin{document}
\title{Theoretical model for the Seebeck coefficient in superlattice materials\\
with energy relaxation}
\author{Vassilios Vargiamidis}
\email{V.Vargiamidis@warwick.ac.uk}
\affiliation{School of Engineering, University of Warwick, Coventry, CV4 7AL, UK}%
\author{Mischa Thesberg}
\affiliation{Institute for Microelectronics, Technical University of Vienna, Vienna, A-1040, Austria}%
\author{Neophytos Neophytou}
\affiliation{School of Engineering, University of Warwick, Coventry, CV4 7AL, UK}%

\begin{abstract}
We present an analytical model for the Seebeck coefficient $S$ of superlattice materials that explicitly takes into account the energy relaxation due to electron-optical phonon (e-ph) scattering. In such materials the Seebeck coefficient is not only determined by the bulk Seebeck values of the materials but, in addition, is dependent on the energy relaxation process of charge carriers as they propagate from the less-conductive barrier region into the more-conductive well region. We calculate $S$ as a function of the well size $d$, where carrier energy becomes increasingly relaxed within the well for $d > \lambda_E$, where $\lambda_E$ is the energy relaxation length. We validate the model against more advanced quantum transport simulations based on the nonequilibrium Green's function (NEGF) method and also with experiment, and we find very good agreement. In the case in which no energy relaxation is taken into account the results deviate substantially from the NEGF results. The model also yields accurate results with only a small deviation (up to $\sim 3\%$) when varying the optical phonon energy $\hbar \omega$ or the e-ph coupling strength $D_\text{0}$, physical parameters that would determine $\lambda_E$. As a first order approximation, the model is valid for nanocomposite materials and it could prove useful in the identification of material combinations and in the estimation of ideal sizes in the design of nano-engineered thermoelectric materials with enhanced power factor performance.
\end{abstract}
 \maketitle

\section{Introduction}

When a temperature gradient is applied in a solid material with free electronic carriers, a voltage gradient arises as carriers migrate from the hot side to the cold side. The strength of this thermoelectric effect is quantified by the Seebeck coefficient $S$, which is defined as the ratio of the voltage difference $\Delta V$ to the temperature difference $\Delta T$. The absolute value of $S$ is referred to as the thermopower.

The Seebeck coefficient is central to the performance of a thermoelectric (TE) material, which is quantified by its TE figure of merit $ZT = \sigma S^2 T / \kappa$, where $\sigma$ is the electronic conductivity, $T$ is the temperature, and $\kappa$ is the thermal conductivity. The product $\sigma S^2$ is known as the power factor ($PF$). Although Bi$_2$Te$_3$ and PbTe are traditionally the most extensively studied TE materials, over the last several years various other materials have been explored with respect to their TE performance, such as transition-metal dichalcogenides \cite{GDing16,HHuang16,WHuang14}, phonon-glass-electron crystals \cite{Beek15nmat}, half-Heuslers \cite{CFu15ncom}, tin selenide \cite{Biswas12}, etc. Most of these materials exhibit $ZT$ above $1$, primarily due to the reduction of their thermal conductivity \cite{Beretta19}.

Superlattices and nanocomposite materials are also currently being explored aiming to achieve even higher TE performance \cite{Mizuno15,Thes15JAP,Thes16JAP,MZhou17,Priy18}. This is due to two reasons. First, they usually cause reduction of the phonon thermal conductivity to ultra-low values as a result of extensive phonon-boundary scattering  \cite{Mizuno15}. In fact, this is considered as one of the most effective ways to enhance TE performance. Second, such nanostructures quite often also cause increase in the Seebeck coefficient \cite{Venk96,Ishida09,Zeng07,Bian07}, and interestingly, in some cases cause increase in the $PF$ as well \cite{Neo13,Vargiam19,NeoJAP13}.

The design of superlattice TE materials requires extensive theoretical and computational modeling. For this purpose, several methods have been employed previously. Some of these methods adopt semi-classical approaches using the Boltzmann transport equation (BTE) where the effects of grain boundaries are treated as a scattering mechanism with some relaxation time \cite{Popescu09,CBera10,MBart01}. However, in order to capture key aspects of the physics - such as tunneling, non-equilibrium carrier relaxation and confinement - in a single setting (especially as the material feature sizes shrink to the nanoscale), the use of a quantum transport method such as the nonequilibrium Green's function (NEGF) method \cite{Koswatta07,Anantram08} is necessary. However, these methods are either complex or time consuming, or both. On the other hand, it is important to be able to determine fast and relatively accurately the Seebeck coefficient of superlattice (SL) materials using simpler models. This can be especially useful in experimental settings, and will guide nanostructured designs that will allow for high Seebeck coefficients, potentially high $PF$s as well.

Simplified models to describe the Seebeck coefficient in superlattices (and nanocomposites) exist and are widely used in the literature. The simplest way is to describe the overall Seebeck coefficient as the weighted average of the Seebeck coefficients of the well and the barrier regions, with the weighting factor being the length of each region \cite{Zeng07}. In an additional step, in order to satisfy the continuity of heat transfer, the individual components are also weighted not only by the length of the regions, but also by the inverse of their thermal conductivity \cite{Kim11}. The individual coefficients are usually obtained from the Boltzmann transport formalism separately for each region. Other works, on the other hand, use the energy dependencies of the coefficients from the BTE and by assuming thermionic emission over the potential barriers \cite{Ishida09,Nard15}. A phenomenological model has also been proposed \cite{Popescu09} for the calculation of the Seebeck coefficient of nanocomposites where interface potential barriers due to grains have been included and the effect of various scattering mechanisms was examined. In more elaborate cases, wave solutions of electronic transport are employed, which account for the formation of mini-bands as well, which are then included in transport \cite{Bian07},although such mini-bands would be weakened in the presence of electron-phonon interaction \cite{Thes16JAP}. In principle, however, there is an intermediate region, where the Seebeck coefficient transits from the barrier into the well and vise-versa, as electrons relax their energy (or gain energy) to go from one region to the other. In structures where the energy relaxation mean-free-path is comparable to the well size, this region becomes important. In fact, we have shown in the past that it is the existence of this region that allows for signiﬁcant power factor improvements in SLs and nanocomposites \cite{Thes16JAP,Neo13,Kim11}. Thus, this region needs to be properly described in compact models that apply to the new generation nanocomposite TE materials, and currently, no compact model exists (despite the importance of it being evident in large scale simulations \cite{Thes15JAP,Thes16JAP,Vargiam19}).

In this paper, we develop a simple analytical model for the Seebeck coefficient of a channel with embedded SL barriers for energy filtering, which mimics either a SL or a nanocomposite to first order approximation, or any material in which carrier transport alternates between potential barriers and wells. Using the average energy of the current flow, and taking the energy relaxation length $\lambda_\text{E}$ as calculated from NEGF we derive an expression for $S$. The relaxation length is generally used to describe the relaxation of the carrier energy along the transport direction due to phonon emission, and is therefore a measure of the distance that the relaxation process occurs. For carriers flowing over barriers and relaxing into wells, it essentially denotes the region where the individual attributes of conductivity and Seebeck coefficient intermix. It is directly connected with the more familiar energy relaxation \emph{time}, $\tau_\text{E}$, which can be calculated for different scattering mechanisms \cite{Lundstrom} and is known for many materials \cite{Selberherr}. 

We present results for the case in which the Fermi level $E_{\text{F}}$ is $\approx k_{\text{B}} T$ below the barrier height, and also for the case where $E_{\text{F}} = V_\text{B}$. We find that the results for $S$ as a function of well size $d$ are in very good agreement with the corresponding results of NEGF. Further, in NEGF we alter the optical phonon energy $\hbar \omega$ and e-ph coupling strength $D_{\text{0}}$, which are the physical parameters that affect the energy relaxation length $\lambda_E$. For all extracted $\lambda_E$'s that we consider, the model predicts accurately the dependence of $S$ on these physical parameters.
\begin{figure}[t]
\vspace{-0.1in}
\hspace*{-0cm}
\includegraphics[width=8.7cm,height=9.1cm]{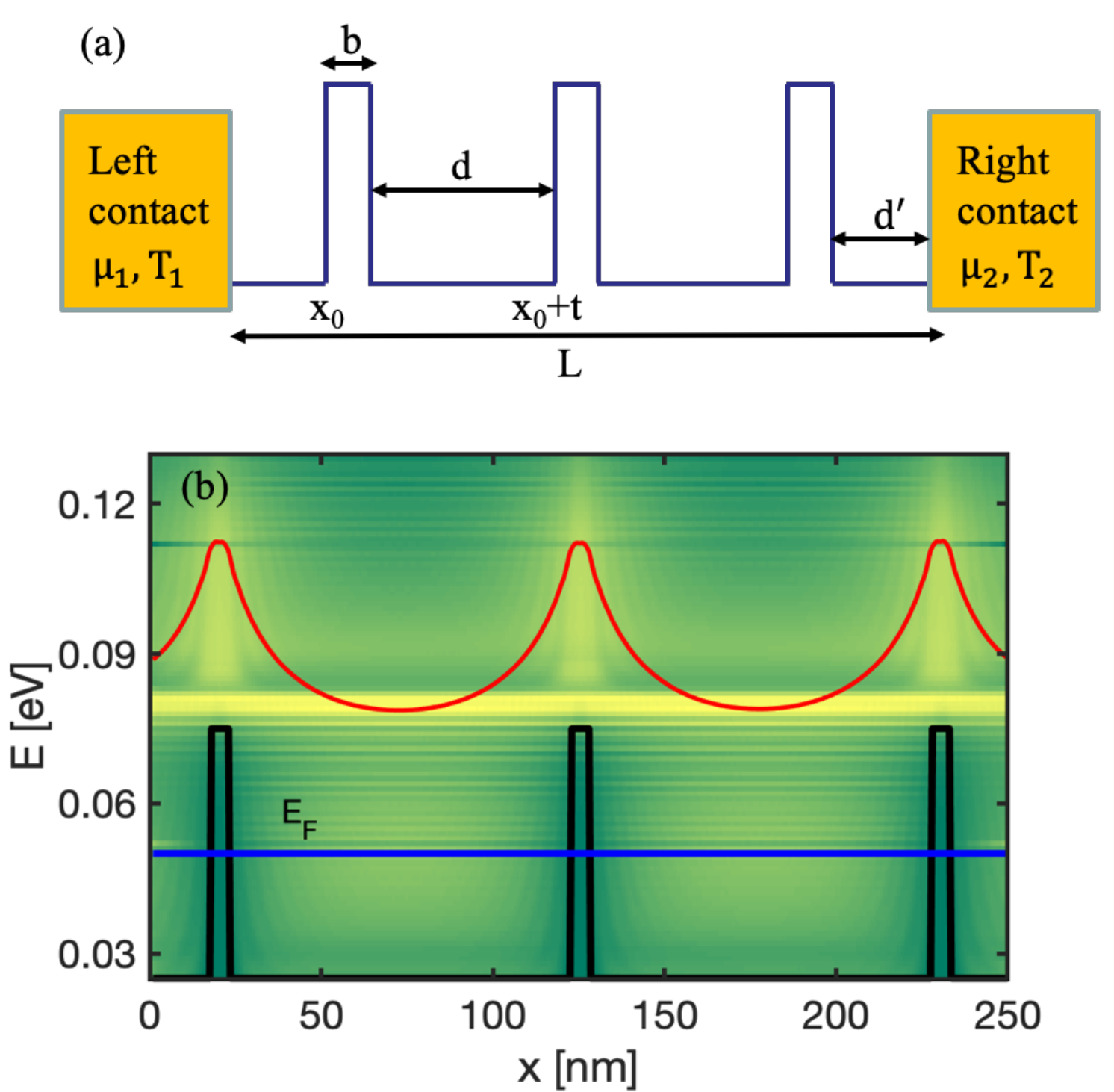}\\
\vspace*{0.0cm} \caption{\label{fig1} (Colour online)
(a) Schematic representation of a superlattice nanostructure. A channel of length $L$ is connected to ideal reservoirs (left and right contacts) with chemical potentials and temperatures $\mu_1$, $T_1$ and $\mu_2$, $T_2$, respectively. The well size is $d$ and the barrier thickness is $b$. We also define $t = b + d$ (see Appendix A). (b) Average energy of the current flow $\langle E(x) \rangle$ as defined in Eq.~(\ref{eq5}) along the channel with SL barriers calculated with NEGF. The black lines represent the potential barriers and the blue line represents the position of the Fermi level $E_\text{F}$. The color map indicates the current flow $I(E,x)$, with yellow indicating high, and green low current density.
}
\end{figure}

The rest of the paper is organized as follows. In Sec.~II we derive the model for the Seebeck coefficient. Then, in Sec.~III, we present, analyse, and validate the results with those from NEGF. In Sec. IV we validate the results with an experiment, while in Sec. V we summarize and conclude.
                                                                                
\section{Seebeck coefficient}

\subsection{Model without energy relaxation}

We consider a nanostructure composed of two different materials in which charge carriers propagate through low and high energy regions, or potential wells and barriers and abstract such a material to that of a one-dimensional (1D)-like system of potential barriers. A schematic representation of such a SL nanostructure is shown in Fig.~1(a). A channel of length $L$ is connected to two contacts (left and right), which are ideal reservoirs in equilibrium with $\mu_1$, $T_1$ and $\mu_2$, $T_2$ their chemical potentials and temperatures, respectively, while $b$ and $d$ are the barrier thickness and well length, respectively. 

Note that, although the system is conceptualized as a SL and only one direction of transport is considered, the final analytical model that we derive below depends only on bulk Seebeck coefficients and energy relaxation lengths of the constituent materials. Thus, the final model is considered to be agnostic to issues of dimensionality and valid for all dimensional structures. Furthermore, we also argue that this model should, on average, be valid for nanocomposite materials. This is because the primary conceptual difference between nanocomposites and SLs is that in a nanocomposite the barriers only have an average spacing of $d$ drawn from a statistical distribution, rather than the fixed rigid spacing of a superlattice. However, since the central crux of the model rests on the dominant effect of carrier relaxation physics, we would expect any coherent phenomena related to quantum reflections, resonances, etc., whose existence separates the nanocomposite and SL cases, to be negligible regardless.

The most commonly used model describes the total Seebeck coefficient of such a system, $S_{\text{sys}}$, as a combination of the Seebeck coefficients of the well and barrier regions, $S_{\text{W}}$ and $S_{\text{B}}$, respectively. This is derived from \cite{Kim11}
\begin{equation}
S_{\text{sys}} = \frac{1}{\Delta T} \int_{0}^{L} S(x) \left( \frac{d T_L}{d x} \right) dx ,
\label{eq1}%
\end{equation}
where $S(x)$ is the local Seebeck coefficient, and $\Delta T$ is the lattice temperature difference along the channel. Note that this expression, strictly speaking, depends on the lattice temperature, $T_\text{L}$, however here we only consider the temperature of carrier flow, $T$.  We take $T_\text{L} = T$, a point justified in Refs.~\cite{Kim11,Kim12} since optical phonon scattering plays the dominant role in energy relaxation (encouraging equilibrium with the phonon bath). Through Fourier's Law, we express the temperature gradient in a barrier (well) region as $\left( d T / d x \right)_{\text{B(W)}} = J / \kappa_{\text{B(W)}}$, where $\kappa_{\text{B}}$ and $\kappa_{\text{W}}$ are the thermal conductivities in the barrier and well regions, respectively, and $J$ is the heat flux. Also, we express $\Delta T$ as
\begin{equation}
\Delta T = \left( \frac{J}{\kappa_{\text{B}}} \right) L_{\text{B}} + \frac{J}{\kappa_{\text{W}}} \left( L_{\text{W}} +L_{\text{W}}^{\prime} \right) ,
\label{eq2}%
\end{equation}
where $L_{\text{B}} = n b$ is the sum of all barrier thicknesses with $n$ the total number of barriers, $L_{\text{W}} = (n-1) d$ is the sum of all well lengths, and $L_{\text{W}}^{\prime} = 2 d^{\prime}$ is the total length of the two wells at the ends of the channel. Note that $x_{\text{0}} = d^{\prime}$ [see Fig.~1(a)]. The reason why the terminating regions are treated separately is to allow direct comparison with NEGF simulation later on, where this is necessary. Using Eq.~(\ref{eq2}), we can express Eq.~(\ref{eq1}) as
\begin{equation}
S_{\text{sys}}^{\text{\text{no rel}}} = \frac{\left( S_{\text{B}} L_{\text{B}} / \kappa_{\text{B}} \right) + ( S_{\text{W}} \tilde{L}_{\text{W}}  / \kappa_{\text{W}} ) }{\left( L_{\text{B}} / \kappa_{\text{B}} \right) + ( \tilde{L}_{\text{W}} / \kappa_{\text{W}} )} ,
\label{eq3}%
\end{equation}
where $\tilde{L}_{\text{W}} = L_{\text{W}} + L_{\text{W}}^{\prime}$.

Although Eq.~(\ref{eq3}) describes well the composite Seebeck coefficient in macroscale materials, when the feature sizes of the composite phases are scaled below a few tens of nanometers, this model is inadequate. The reason is that in the vicinity of the materials' interfaces, the Seebeck coefficient does not abruptly change from $S_{\text{B}}$ to $S_{\text{W}}$, but carriers have to gradually relax their energy (and also their momentum) to the value imposed by the equilibrium conditions of each material. In fact, this takes place within a distance determined by the energy relaxation length $\lambda_{\text{E}}$ \cite{Kim11,Kim12}, see Fig.~1(b). It is important to note here that any possible $PF$ improvement in such materials originates from the intermixing of the high Seebeck coefficient of barrier material $S_{\text{B}}$ with the high conductivity of the well material $\sigma_{\text{W}}$, thus making these regions very important in composite nanostructures. We emphasize that, just by considering separately the individual (bulk) Seebeck coefficients of the two regions $S_\text{B}$ and $S_\text{W}$, it is not easy to achieve $PF$ improvements compared to the maximum of the two $PF$s of the individual barrier or well. 

In Sec.~IIB, we develop a simple and relatively accurate model for the Seebeck coefficient of SL materials, taking into account the energy relaxation process due to e-ph scattering. We validate the model against the results from NEGF. In the NEGF simulations we consider only electron-optical phonon scattering mediated through the e-ph coupling strength $D_{\text{0}}$, which is the mechanism most responsible for energy relaxation.

\subsection{Model with energy relaxation} 

In order to derive an analytical model for the Seebeck coefficient of a nanocomposite system consisting of potential barriers and wells as shown in Fig.~1(a), we assume that the charge carriers are fully relaxed in the barrier regions, but in the well regions the carriers undergo a relaxation process, which is quantified by $\lambda_E$. Apart from the optical phonon energy $\hbar \omega$ and the deformation potential $D_\text{0}$, a key parameter that influences the relaxation process is the well size $d$, as discussed below. 

The $x$ dependent (local) Seebeck coefficient is given as
\begin{equation}
S(x) = \frac{\langle E(x) \rangle - E_{\text{F}}}{q T} ,
\label{eq4}%
\end{equation}
where $q$ is the carrier charge ($q = - |e|$ for electrons and $q = + |e|$ for holes), $E_\text{F}$ is the Fermi level, and $\langle E(x) \rangle$ is the average energy of the current flow along the $x$ direction (propagation direction), defined as	
\begin{equation}
\langle E ( x ) \rangle = \frac{1}{J} \int_{E=E_c}^{E=\infty} I (E, x) E dE  .
\label{eq5}%
\end{equation}	
In Eq.~(\ref{eq5}), $I(E,x)$ is the energy and position resolved current, while $J = \int I(E,x) dE=\text{constant}$. Note that even though the current is constant along the channel at each cross section, its energy is not constant, i.e., the charge carriers can gain or lose energy as they propagate. This happens in the presence of inelastic scattering (optical phonons). We emphasize that knowledge of $\langle E(x) \rangle$ in Eq.~(\ref{eq4}) allows one to determine the Seebeck coefficient $S(x)$ and vice versa, regardless the complexities of the nanostructure.
\begin{figure}[t]
\vspace{-0.1in}
\hspace*{-0.2cm}
\includegraphics[width=9.4cm,height=11.4cm]{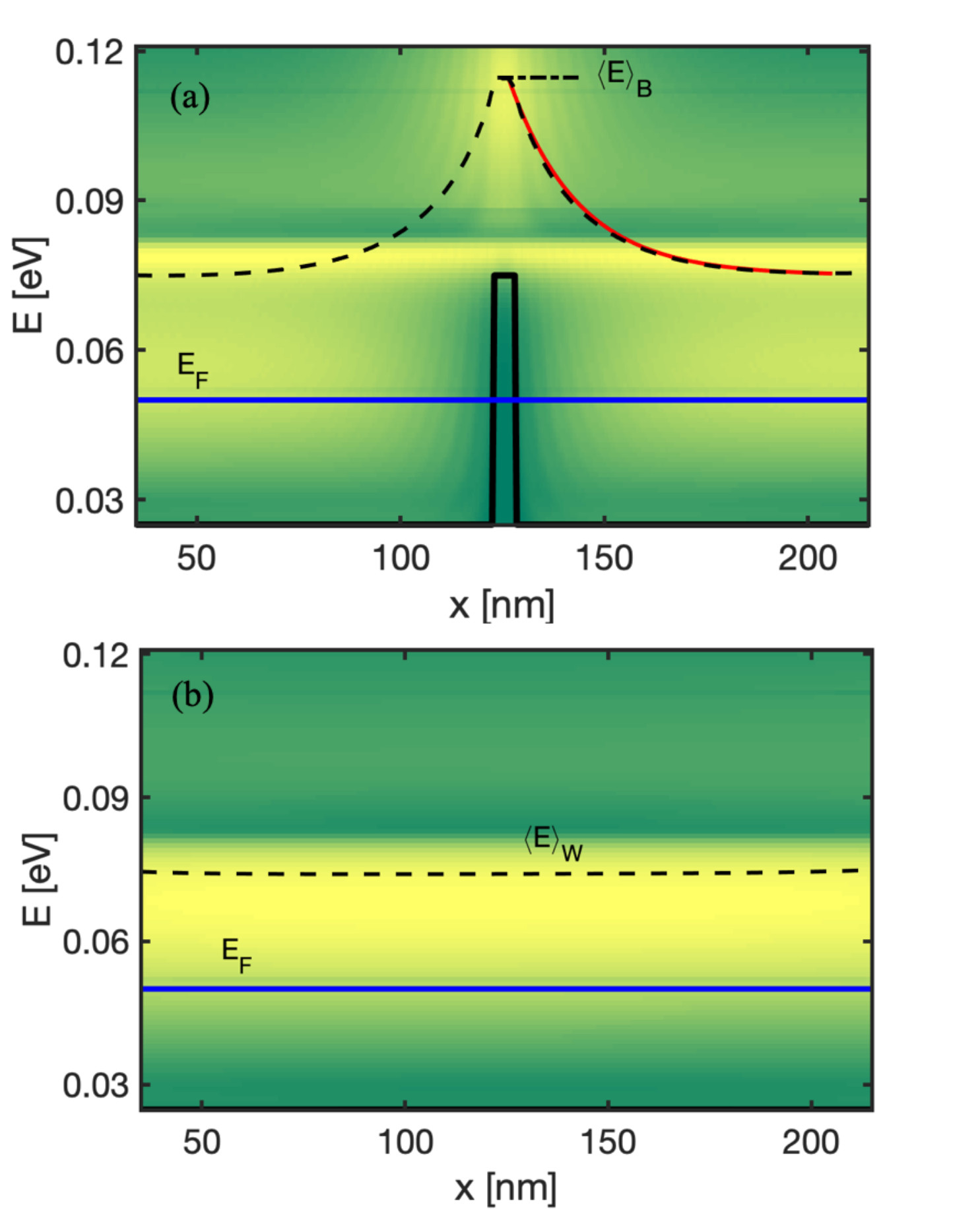}\\
\vspace*{0cm} \caption{\label{fig2} (Colour online)
Average energy of the current flow $\langle E(x) \rangle$ (dashed, black lines) along the channel with (a) a single barrier, and (b) no barrier and $E_\text{F} =0.05$eV, calculated with NEGF. The solid (red) line is fitting of Eq.~(\ref{eqA5}) in order to extract the energy relaxation $\lambda_{\text{E}}$ (see Sec.~III). $\langle E \rangle_{\text{B}}$ and $\langle E \rangle_{\text{W}}$ are the average energies on the barrier, and in the well under equilibrium, respectively. The color map indicates the current flow $I(E,x)$, with yellow indicating high, and green low current density.
}
\end{figure}

We substitute now $S(x)$ from Eq.~(\ref{eq4}) into Eq.~(\ref{eq1}) and we obtain
\begin{equation}
S_{\text{sys}} = \frac{1}{q T \Delta T} \int_0^L \left( \langle E(x) \rangle - E_F \right) \left( \frac{d T_{\text{L}}}{d x} \right) dx  .
\label{eq6}%
\end{equation}	
In Eq.~(\ref{eq6}), in order to make contact with NEGF simulation, it is necessary to consider $\langle E(x) \rangle$ in the well regions as being different from that in the wells at the channel ends (i.e., those that are close to the contacts) as shown in Appendix A. Further, in the barrier regions $\langle E(x) \rangle$ is taken to be constant (see Eq.~(\ref{eqA1}) of Appendix A), which turns out to be a good approximation. Accordingly, Eq.~(\ref{eq6}) can be expressed as a sum of four terms, each of which pertains to different region, i.e., well regions, wells at the left and right channel ends, and barrier regions, $\tilde{s}_{\text{W}}$, $\tilde{s}_{\text{L}}$, $\tilde{s}_{\text{R}}$, and $\tilde{s}_{\text{B}}$, respectively. The total Seebeck coefficient can then be expressed as
\begin{equation}
S_{\text{sys}} = \tilde{s}_{\text{L}} + n \tilde{s}_{\text{B}} +(n-1) \tilde{s}_{\text{W}} + \tilde{s}_{\text{R}} ,
\label{eq7}%
\end{equation}	
where $n$ is the total number of barriers in the channel. Each one of the four terms in Eq.~(\ref{eq7}) is of the same form as Eq.~(\ref{eq6}), but each one with different form of average energy $\langle E(x) \rangle$ and with different integration region.

Starting with $\tilde{s}_{\text{B}}$, we use Eq.~(\ref{eqA1}) for a single barrier, i.e.,
\begin{equation}
\langle E(x) \rangle = \langle E \rangle_{\text{B}}  ,
\label{eq8}%
\end{equation}	
where $\langle E \rangle_{\text{B}}$ is the average energy on the top of a barrier [see Fig.~2(a)]. Thus,
\begin{equation}
\tilde{s}_{\text{B}} = \left( \frac{b}{\Delta T} \right) \frac{J}{\kappa_{\text{B}}} S_{\text{B}} ,
\label{eq9}%
\end{equation}	
where
\begin{equation}
S_{\text{B}} = \frac{\langle E \rangle_{\text{B}} - E_{\text{F}}}{q T} .
\label{eq10}%
\end{equation}	
In order to evaluate $S_{\text{B}}$ in Eq.~(\ref{eq10}), we use NEGF to evaluate first the average energy $\langle E \rangle_{\text{B}}$ on top of a single barrier placed in the channel, as shown in Fig.~2(a). However, in general the value of $S_{\text{B}}$ could be extracted from bulk experimental values. As it turns out $S_{\text{B}}$ is actually one of three parameters needed to determine the total Seebeck coefficient $S_{\text{sys}}$.

For the calculation of $\tilde{s}_{\text{W}}$ we use Eq.~(\ref{eqA3}) for a single well (see rationale in the Appendix), as also seen in Ref.~\cite{Kim12},
\begin{eqnarray}
\nonumber \hspace{-0.5cm} \langle E(x) \rangle = \left( \langle E \rangle_{\text{B}} - \langle E \rangle_{\text{W}} \right) 
\\* &&\hspace*{-1.74in} \times \left( e^{-(x - x_0 -b) / \lambda_{\text{E}}} + e^{(x-x_0-t)/\lambda_{\text{E}}} - e^{-d / \lambda_{\text{E}}} \right) + \langle E \rangle_{\text{W}}  ,
\label{eq11}%
\end{eqnarray}
where $\langle E \rangle_{\text{W}}$ is the average energy within the well under equilibrium and $x_0 + b < x <x_0 +  t$. The value of $\langle E \rangle_\text{W}$ can be extracted from NEGF as the average energy of the current flow for a pristine channel (as is done here), or taken from experimental measurement. The result is then that of an "effective" well, as shown in Fig.~2(b). Using Eq.~(\ref{eq11}), we can express $\tilde{s}_W$ in the form
\begin{equation}
\tilde{s}_{\text{W}} = \left( \frac{d}{\Delta T} \right) \frac{J}{\kappa_{\text{W}}} S_{\text{W-relax}}  ,
\label{eq12}%
\end{equation}	
where $S_{\text{W-relax}}$ is obtained using Eq.~(\ref{eq6}) and the form of $\langle E(x) \rangle$ given in Eq.~(\ref{eq11}). It is given by
\begin{eqnarray}
\nonumber \hspace{-0.9cm} S_{\text{W-relax}} = S_{\text{W}} + \left( S_{\text{B}} - S_{\text{W}} \right)  
\\* &&\hspace*{-1.1in} \times \left( \frac{2 \lambda_{\text{E}}}{d} \right) \left[ 1 - e^{-d / \lambda_{\text{E}}} \left( 1 + \frac{d}{2 \lambda_{\text{E}}} \right)  \right] .
\label{eq13}%
\end{eqnarray}
In Eq.~(\ref{eq13}) we made use of Eq.~(\ref{eq10}) for $S_{\text{B}}$ and the corresponding relation for $S_{\text{W}}$. The value of $S_\text{W}$ is obtained from $\langle E \rangle_\text{W}$ of the pristine channel shown in Fig.~2(b). Note that in the two limits $\lambda_{\text{E}} \rightarrow 0$ and $\lambda_{\text{E}} \gg d$, Eq.~(\ref{eq13}) yields $S_{\text{W-relax}} \rightarrow S_{\text{W}}$ and $S_{\text{W-relax}} \rightarrow S_{\text{B}}$, respectively, i.e., the corresponding bulk values of the Seebeck coefficient in the well and barrier regions, as expected. A similar calculation for $\tilde{s}_{\text{R}}$ using Eq.~(\ref{eqA5}) yields
\begin{equation}
\tilde{s}_{\text{R}} = \left( \frac{d^{\prime}}{\Delta T} \right) \frac{J}{\kappa_{\text{W}}} S_{\text{W-relax}}^{\prime}  ,
\label{eq14}%
\end{equation}	
where
\begin{equation}
S_{\text{W-relax}}^{\prime} = S_{\text{W}} + \left( S_{\text{B}} - S_{\text{W}} \right) \left( \frac{\lambda_{\text{E}}}{d^{\prime}} \right) \left( 1 - e^{-d^{\prime} / \lambda_{\text{E}}} \right)   .
\label{eq15}%
\end{equation}	
Note that $d^{\prime} = x_{\text{0}}$ (see Fig.~1(a)). The contribution of $\tilde{s}_{\text{L}}$ is identical to that of $\tilde{s}_{\text{R}}$, i.e., $\tilde{s}_{\text{L}} = \tilde{s}_{\text{R}}$. Thus, finally, from Eq.~(\ref{eq7}) the total Seebeck coefficient of the system takes the form
\begin{eqnarray}
\nonumber \hspace{-0cm} S_{\text{sys}}=  
\\* &&\hspace*{-0.5in} \nonumber \frac{\left( L_{\text{B}} S_{\text{B}}  / \kappa_{\text{B}} \right) + ( L_{\text{W}} S_{\text{W-relax}}  / \kappa_{\text{W}} ) +( L_{\text{W}}^{\prime} S_{\text{W-relax}}^{\prime}  / \kappa_{\text{W}} ) }{\left( L_{\text{B}} / \kappa_{\text{B}} \right) + ( \tilde{L}_{\text{W}} / \kappa_{\text{W}} ) }  .\\
\label{eq16}%
\end{eqnarray}
In Eq.~(\ref{eq16}) we notice that - aside from the geometric factors of $L_\text{W}$, $L_\text{B}$, etc. - the Seebeck coefficient of the SL material is determined in terms of only three parameters; namely, the bulk Seebeck coefficients of the barrier and well materials $S_{\text{B}}$ and $S_{\text{W}}$, respectively, and the energy relaxation length $\lambda_{\text{E}}$. The difference between Eq.~(\ref{eq3}) and Eq.~(\ref{eq16}) is the presence of energy relaxation in the latter, which is a result of electron-optical phonon scattering. In fact, the energy relaxation process leads to the partial extension of $S_{\text{B}}$ into the well region up to a distance $\approx \lambda_{\text{E}}$ from the barrier. We remark that in a long channel we can neglect the two wells close to the contacts to a good approximation, in which case Eq.~(\ref{eq16}) can be expressed as
\begin{equation}
S_{\text{sys}} \simeq \frac{\left( L_{\text{B}} S_{\text{B}}  / \kappa_{\text{B}} \right) + (  L_{\text{W}} S_{\text{W-relax}}  / \kappa_{\text{W}} ) }{\left( L_{\text{B}} / \kappa_{\text{B}} \right) + ( L_{\text{W}} / \kappa_{\text{W}} ) }    .
\label{eq17}%
\end{equation}	

Note that the presence of barriers causes increase of the average energy of the current and a consequent increase in the Seebeck coefficient, which is due to the energy filtering provided by the barriers \cite{NeoJAP13}. On the other hand, the energy relaxation process of charge carriers in the well region causes electrons to propagate at lower energy states, which leads to reduction of the Seebeck coefficient. However, at the edge of a well and immediately after the barrier the Seebeck coefficient remains close to its highest value, i.e., close to $S_\text{B}$. The conductivity mean-free-path of the carriers of these spatial regions can still be high as carriers propagate at higher velocity states compared to the relaxed well states, resulting in an increase of the $PF$ for suitable well sizes \cite{Neo13,Neop14,ThesPRB17}. This increase originates from these energy non-relaxed regions, which emphasizes the need for them to be captured accurately.

\section{Validation of model with NEGF}

We compare now the analytical model for the Seebeck coefficient with simulation results from 1D NEGF. However, we re-iterate that NEGF simulation plays the role here of validation, and the reduced dimensionality considered is the result of computational necessity on the part of NEGF simulation, but the analytical model itself is expected to be dimensionally agnostic. We consider a channel of length $L = 250$nm, an initial number of twenty four rectangular barriers ($n = 24$) with spacing $d = 4$nm between them, and each one of thickness $b = 5$nm and height $V_{\text{B}} = 0.05$eV. For simplicity we consider equal thermal conductivities for the barrier and well materials, i.e., $\kappa_{\text{B}} = \kappa_{\text{W}}$. For the NEGF simulations we assume the same parameter values and a channel with arbitrarily small width $W = 3$nm to help with convergence which can be difficult in a truly 1D structure with such intense optical phonon scattering. The small width of the channel gives rise to an upward shift of the subband energies by an amount $0.025$eV, resulting in an effective barrier height of $V_{\text{B}} = 0.075$eV. The channel with the effective barrier height is as shown in Fig.~3(a).
\begin{figure}[t]
\vspace{-0.1in}
\hspace*{-0.2cm}
\includegraphics[width=9.2cm,height=11.6cm]{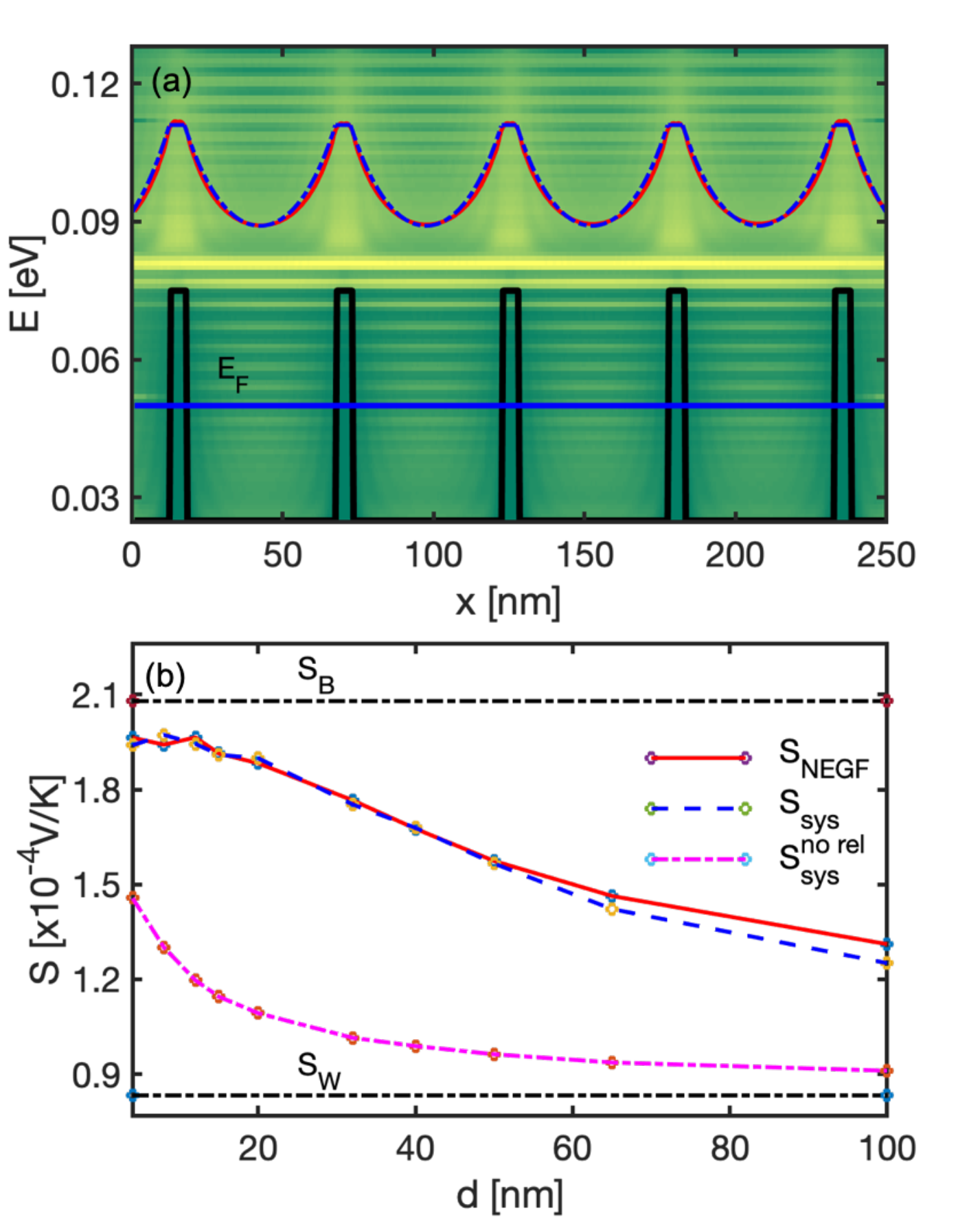}\\
\vspace*{0cm} \caption{\label{fig3} (Colour online)
(a) Average energy of the current flow $\langle E(x) \rangle$ along the channel with five barriers calculated from Eq.~(\ref{eq5}) using NEGF (solid, red line) and from the analytical result Eq.~(\ref{eqA3}) (dashed, blue line) for which $d = 50$nm and $\lambda_\text{E} = 16.5$nm.The color map indicates the current flow $I(E,x)$, with yellow indicating high, and green low current density. (b) Seebeck coefficient vs well size $d$ calculated from (i) NEGF (solid, red line), (ii) model with energy relaxation (dashed, blue line), and (iii) model without energy relaxation (dashed-dotted, magenta line).
}
\end{figure}
\begin{figure}[t]
\vspace{-0.1in}
\hspace*{-0.15cm}
\includegraphics[width=9.4cm,height=11.6cm]{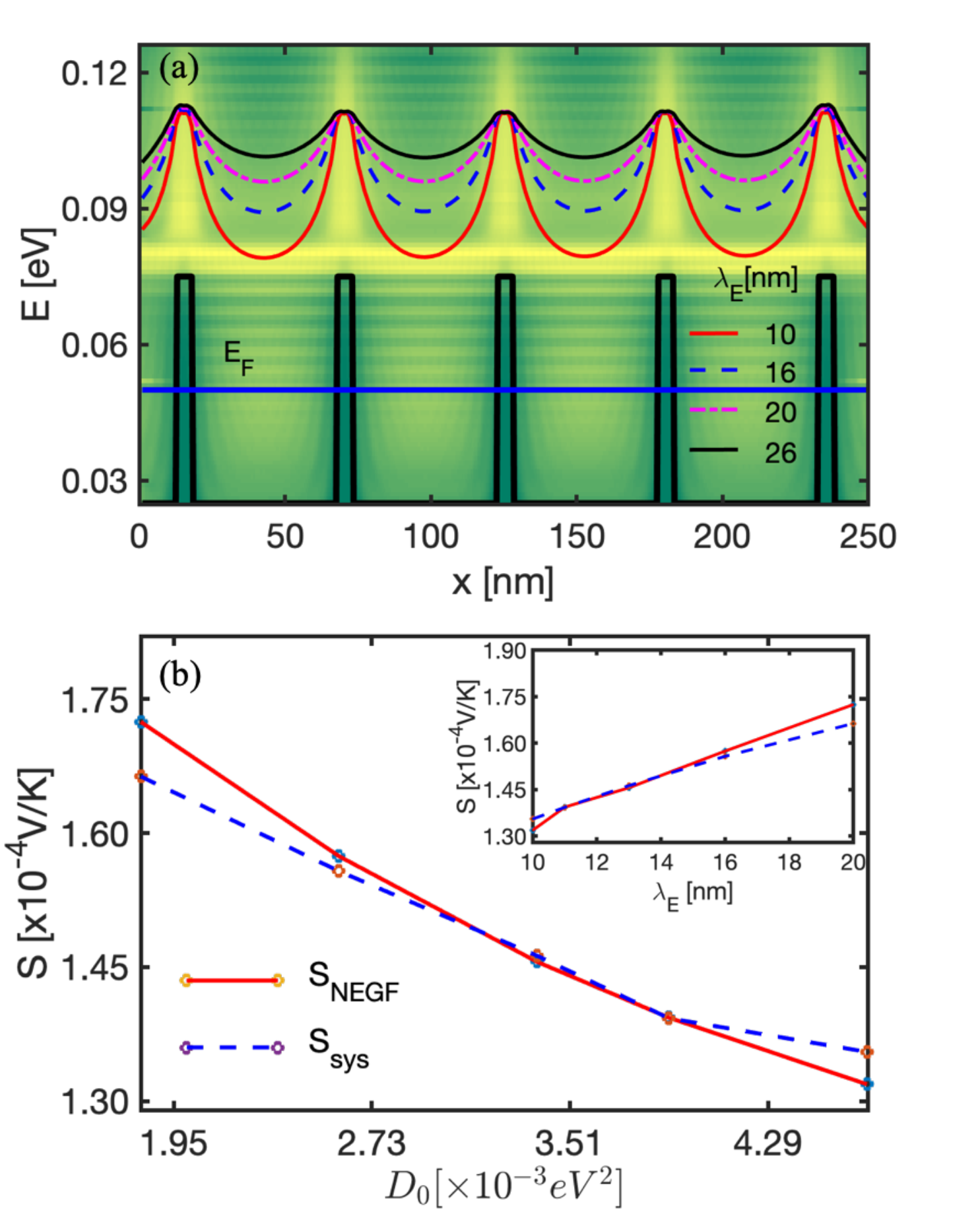}\\
\vspace*{0cm} \caption{\label{fig4} (Colour online)
(a) Average energy of the current flow $\langle E(x) \rangle$ along a channel with five barriers calculated from Eq.~(\ref{eq5}) using NEGF for increasing values of $\lambda_\text{E}$ (i.e., decreasing values of the deformation potential $D_\text{0}$). The color map indicates the current flow $I(E,x)$, with yellow indicating high, and green low current density. (b) Seebeck coefficient vs $D_\text{0}$ calculated with NEGF (solid, red line) and with the model (dashed, blue line). Inset: the same but vs $\lambda_\text{E}$.
}
\end{figure}

\subsection{$S$ vs $d$}

In the following we use an e-ph coupling strength $D_{\text{0}} = 0.0026$eV$^2$ and optical phonon energy of $\hbar \omega = 0.06$eV (which is close to the Si value), while we place the Fermi level at $E_{\text{F}} = 0.05$eV. This corresponds to degenerately doped channel, where high $PF$s were observed \cite{Neo13}. The relaxation length is extracted by fitting Eq.~(\ref{eqA5}) [solid red line in Fig.~2(a)] on the NEGF simulation for $\langle E(x) \rangle$ yielding $\lambda_{\text{E}} = 16.5$nm. For these parameter values we also find $\langle E \rangle_{\text{B}} = 0.112$eV and $\langle E \rangle_{\text{W}} = 0.075$eV, as described in the context of Fig.~2. Using Eq.~(\ref{eq10}) and a similar one for $S_\text{W}$ we find $S_\text{B} = 2.06 \times 10^{-4}$V/K and $S_\text{W} = 0.833 \times 10^{-4}$V/K. In Fig.~3(a) we show the average energy of the current flow $\langle E(x) \rangle$ calculated from NEGF (solid red line) and that from the model, i.e., Eqs.~(\ref{eqA1}) and (\ref{eqA3}) (dashed blue line) in the case of five barriers for which $d = 50$nm and $x_{0} = 12.5$nm. We notice that $\langle E(x) \rangle$ plotted from the model describes very well the simulation results and captures all essential features of the current flow, including the relaxation process. We remark here that the decay rate of $\langle E(x) \rangle$ within each well depends on the size of the well compared to the energy relaxation length, as discussed below [see also Eq.~(\ref{eq11})].

In Fig.~3(b) we show the Seebeck coefficient $S_{\text{NEGF}}$ calculated from NEGF (solid, red line) as a function of well size $d$, and the Seebeck coefficient $S_{\text{sys}}$ calculated from the model with energy relaxation Eq.~(\ref{eq16}) (dashed, blue line). The dashed-dotted magenta line shows the Seebeck coefficient $S_{\text{sys}}^{\text{no rel}}$ from the model without energy relaxation Eq.~(\ref{eq3}). $S_{\text{NEGF}}$ is calculated by integrating the average energy of the current flow with respect to the Fermi level when a voltage difference $\Delta V$ is applied at the channel contacts, as shown in Eqs.~(\ref{eq4}) - (\ref{eq6}) (see also Ref.~\cite{Seebeck}). The bulk values of the Seebeck coefficients $S_{\text{B}}$ and $S_{\text{W}}$ are also shown for reference (dashed-dotted, black likes). In these calculations $d$ increases by removing barriers sequentially one at a time while keeping $L$ fixed. Notice that in our model we take into account the finite thickness of the barriers via Eq.~(\ref{eqA1}). Notice that $S_{\text{sys}}$ decreases with increasing $d$ as a consequence of increasing energy relaxation in the well regions, and agrees very well with $S_{\text{NEGF}}$. As $d$ increases, $\langle E(x) \rangle$ gradually relaxes more in the well regions and it approaches $\langle E \rangle_{\text{W}}$ in the middle of each well. However, $d$ should be significantly larger than $\lambda_{\text{E}}$ in order to have full energy relaxation and to achieve the limits $\langle E(x) \rangle \rightarrow \langle E \rangle_\text{W}$ and $S_{\text{sys}} \rightarrow S_{\text{W}}$.

\subsection{$S$ vs $D_\text{0}$}

We compare now the model with the NEGF simulations in the case in which the e-ph coupling strength $D_{\text{0}}$ is varied. Here, again $\hbar \omega = 0.06$eV. In order to illustrate our results we use a channel with five SL barriers. Since the energy relaxation length $\lambda_{\text{E}}$ decreases as $D_\text{0}$ increases, for each value of $D_{\text{0}}$ separately we made fitting of Eq.~(\ref{eqA5}) to the NEGF $\langle E(x) \rangle$ as in Fig.~2(a) and extracted the corresponding values of $\lambda_{\text{E}}$. The values of $\langle E \rangle_\text{B}$ and $\langle E \rangle_\text{W}$ are determined as before. However, as $D_{\text{0}}$ varies, $\langle E \rangle_\text{W}$ does not remain constant and varies slightly. Accordingly, $S_\text{W}$ is determined separately for each value of $D_{\text{0}}$.
\begin{figure}[t]
\vspace{-0.1in}
\hspace*{-0.2cm}
\includegraphics[width=9.2cm,height=11.6cm]{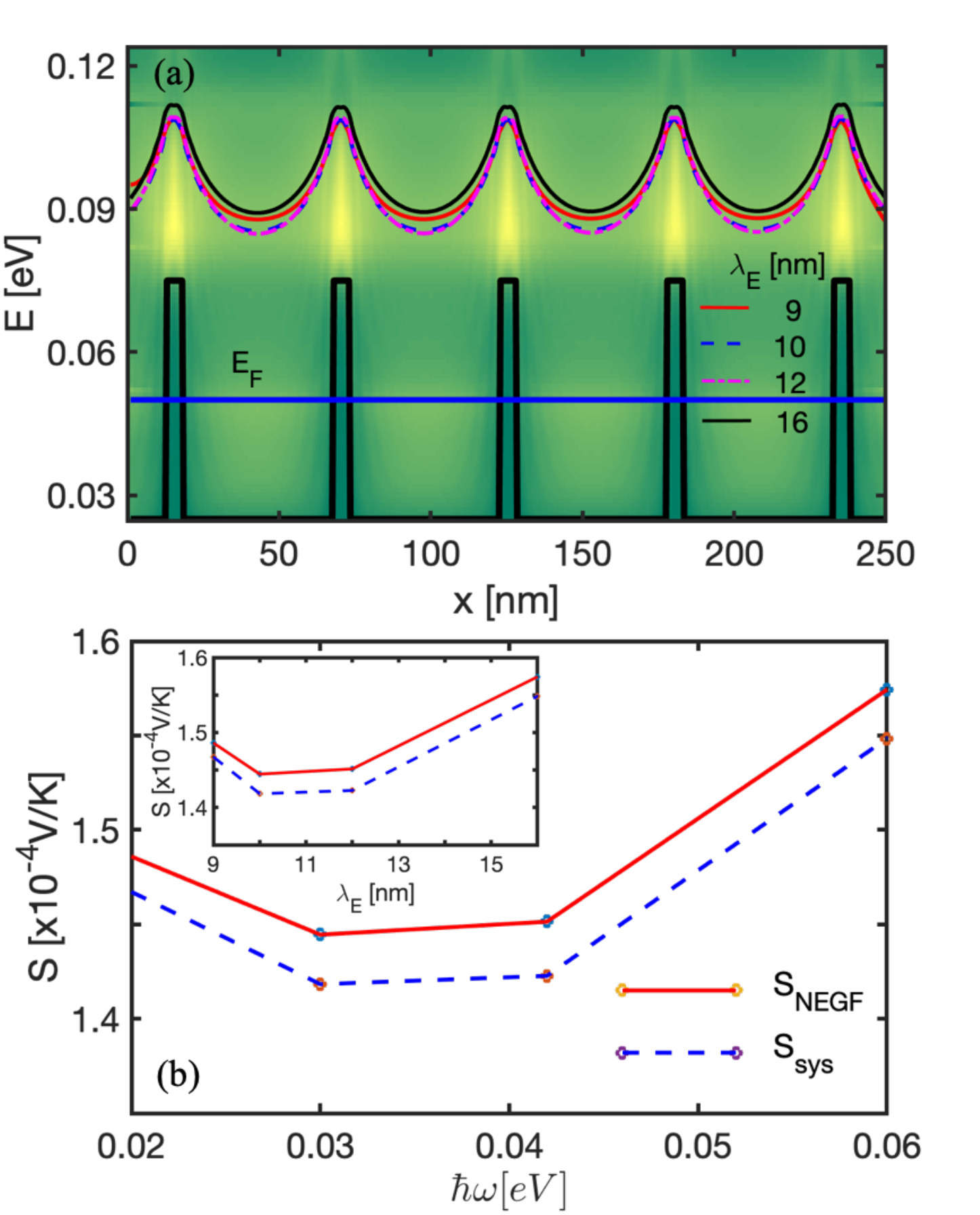}\\
\vspace*{0cm} \caption{\label{fig5} (Colour online)
(a) Average energy of the current flow $\langle E(x) \rangle$ along a channel with five barriers calculated from Eq.~(\ref{eq5}) using NEGF for increasing values of $\lambda_\text{E}$ (i.e., increasing values of the optical phonon energy $\hbar \omega$). The color map indicates the current flow $I(E,x)$, with yellow indicating high, and green low current density. (b) Seebeck coefficient vs $\hbar \omega$ calculated with NEGF (solid, red line) and with the model (dashed, blue line). Inset: the same but vs $\lambda_\text{E}$.
}
\end{figure}

In Fig.~4(a) we show the average energy of the current flow $\langle E(x) \rangle$ calculated with NEGF for increasing values of $\lambda_\text{E}$. Notice the gradually faster decay rate of $\langle E(x) \rangle$ within each well as $\lambda_\text{E}$ decreases (i.e., as $D_{\text{0}}$ increases). The physical origin of this behavior is the enhanced scattering of electrons with optical phonons as $D_\text{0}$ increases, resulting in gradually stronger carrier energy relaxation. Consequently the Seebeck coefficient decreases with increasing $D_\text{0}$, as shown in Fig.~4(b). The solid red line is the NEGF simulation results while the dashed blue line is the result of the model. It can be seen that the agreement between the two results is very good. The inset shows the Seebeck coefficient vs the values of the energy relaxation length $\lambda_\text{E}$ that correspond to the values of $D_\text{0}$ that were used.

\subsection{$S$ vs $\hbar \omega$}

We illustrate now the case in which the optical phonon energy $\hbar \omega$ is varied. Here, we fix $D_\text{0} = 0.0026$eV$^2$ despite the fact that the e-ph coupling strength is $\sim 1 / \omega$ \cite{Lundstrom}, because we intent to investigate the effects of the phonon energy alone, independent of the e-ph coupling. Again we use a channel with five SL barriers. In order to determine the values of $\lambda_\text{E}$ we made fitting of Eq.~(\ref{eqA5}) to the NEGF $\langle E(x) \rangle$ separately for each value of $\hbar \omega$. Also, for each value of $\hbar \omega$ we find the corresponding value of $\langle E \rangle_\text{W}$ as described in the context of Fig.~2(b). In Fig.~5(a) we show the average energy of the current flow $\langle E(x) \rangle$ calculated with NEGF for increasing values of $\lambda_\text{E}$. We notice that for longer energy relaxation lengths, i.e., $\lambda_\text{E} = 16$nm (where $\hbar \omega > V_\text{B} - E_\text{F}$) charge carriers, which travel at energies $\approx k_\text{B} T$ above the barrier height, cannot now easily emit phonons as the final scattering states reside below $E_\text{F}$, and they are almost filled, thus the relaxation rate is lower [see black line in Fig.~5(a)]. In fact, we have performed simulations with even higher $\hbar \omega$, and found that the relaxation rate is suppressed even more. This could be a generic filtering design direction to suppress relaxation in nanostructured materials, by choosing $V_\text{B} - E_\text{F}$ smaller compared to the material's $\hbar \omega$. Thus, the average energy of the current flow increases and as a consequence the Seebeck coefficient also increases. This is shown in Fig.~5(b) where it can also be seen that $S_{\text{sys}}$ agrees very well with $S_{\text{NEGF}}$ to an accuracy of $1$-$2 \%$. We also note that for $\hbar \omega = 0.02$eV the Seebeck coefficient increases slightly. In fact, as $\hbar \omega$ becomes even smaller the energy relaxation gradually diminishes and, in the limit $\hbar \omega \rightarrow 0$, the energy $\langle E(x) \rangle$ should become constant reflecting the absence of optical phonon processes (i.e., resembling the elastic acoustic phonon case or even the ballistic one). The inset shows the Seebeck coefficient vs the values of $\lambda_\text{E}$ that correspond to the values of $\hbar \omega$ on the $x$ axis. 
\begin{figure}[t]
\vspace{-0.1in}
\hspace*{-0.45cm}
\includegraphics[width=9.7cm,height=6.2cm]{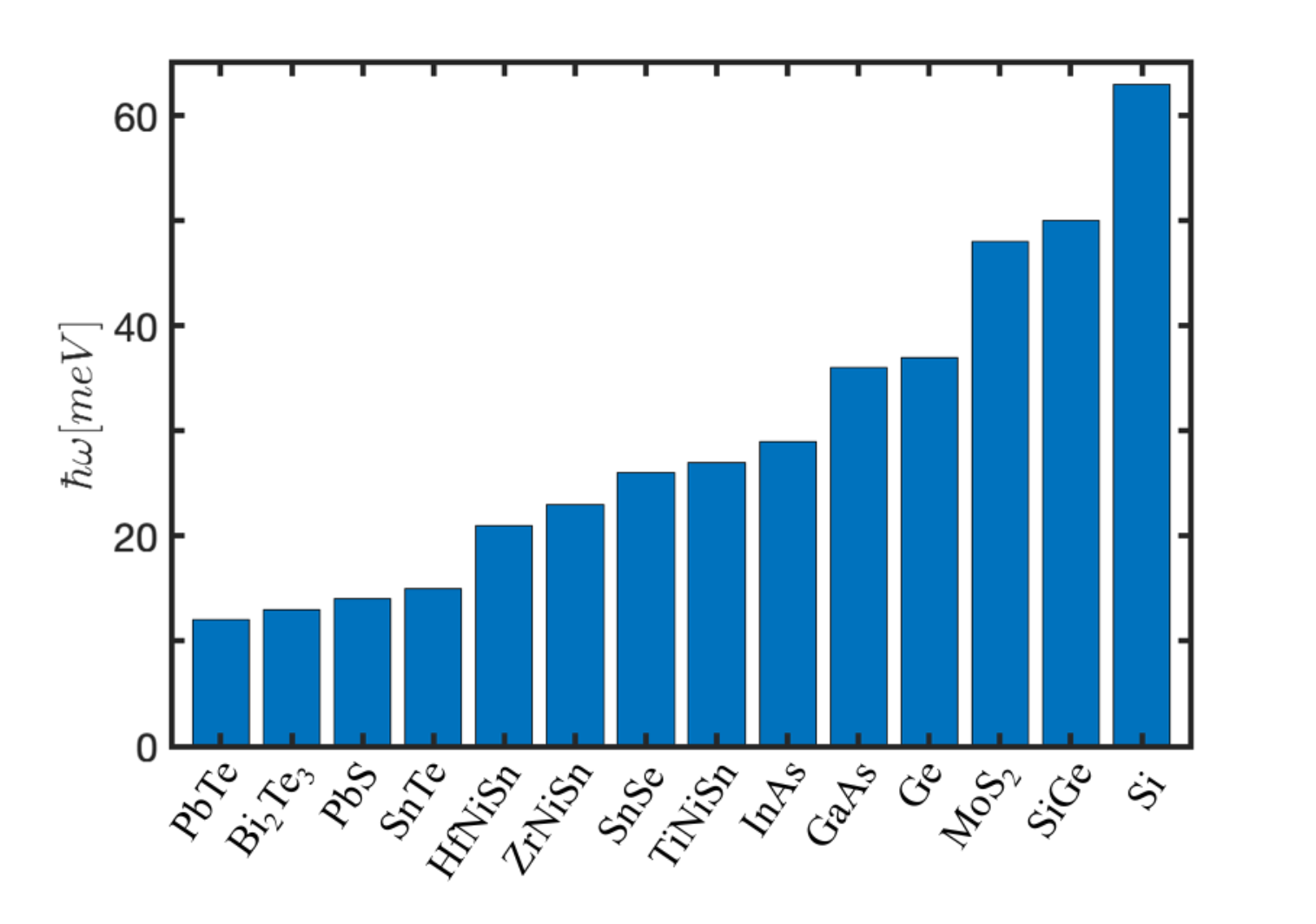}\\
\vspace*{0cm} \caption{\label{fig6} (Colour online)
Optical phonon energies $\hbar \omega$ for common thermoelectric materials.
}
\end{figure}

The values of $\hbar \omega$ for which we plotted the Seebeck coefficient correspond to those of the most common TE materials and semiconductors that are being explored experimentally. The optical phonon energies for some of these materials is shown in Fig.~6, where it is seen that $\hbar \omega$ ranges from $\approx 13$meV for PbTe up to $\approx 63$meV for Si. In passing, we remark that the slightly lower values of $S_{\text{sys}}$ throughout the range of $\hbar \omega$ compared to $S_{\text{NEGF}}$ in Fig.~5(b) are probably due to the fact that $\langle E \rangle_\text{B}$ in the model is constant on the top of each barrier and slightly lower than $\langle E(x) \rangle$ of NEGF.
\begin{figure}[t]
\vspace{-0.1in}
\hspace*{-0.2cm}
\includegraphics[width=9.3cm,height=11.6cm]{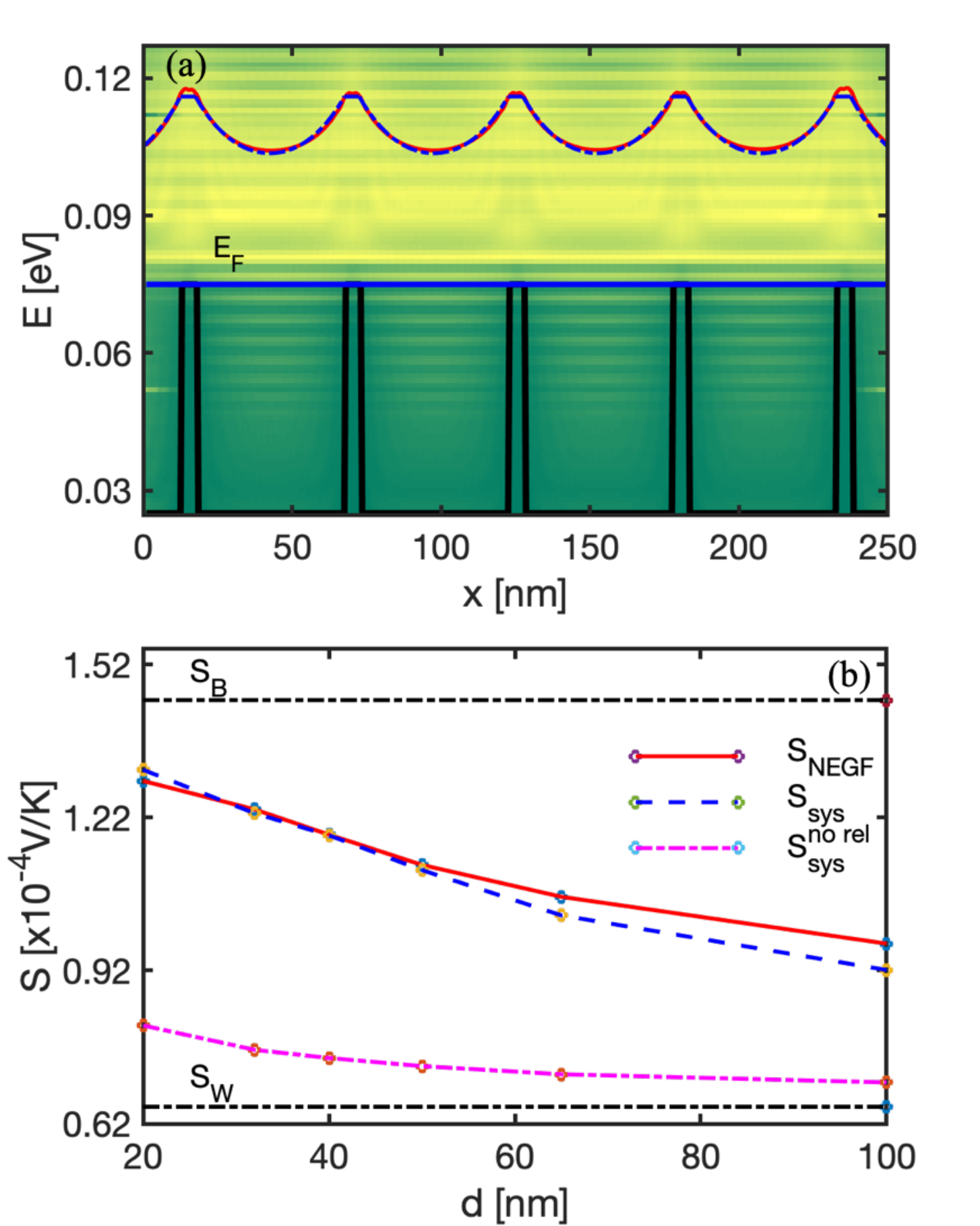}\\
\vspace*{0cm} \caption{\label{fig7} (Colour online)
(a) Average energy of the current flow $\langle E(x) \rangle$ along the channel with five barriers calculated from Eq.~(\ref{eq5}) using NEGF (solid, red line) and from the analytical result Eq.~(\ref{eqA3}) (dashed, blue line) for which $d = 50$nm and $\lambda_\text{E} = 17.5$nm. Here $E_\text{F} = V_\text{B} = 0.075$eV. The color map indicates the current flow $I(E,x)$, with yellow indicating high, and green low current density. (b) Seebeck coefficient vs well size $d$ calculated from (i) NEGF (solid, red line), (ii) model with energy relaxation (dashed, blue line), and (iii) model without energy relaxation (dashed-dotted, magenta line).
}
\end{figure}

\subsection{$S$ vs $d$ ($E_\text{F} = V_\text{B}$)}

In the above analysis the Fermi level was taken at $E_\text{F} = 0.05$eV, i.e., $\approx k_\text{B} T$ below the barrier height. We explore now the regime for which the Fermi level is aligned with the barrier height $V_{\text{B}}$, i.e., $E_\text{F} = V_\text{B} = 0.075$eV, while the rest of the parameters are the same as previously. This case is interesting because it has been shown previously \cite{Vargiam19} that this is the optimal case for $PF$ improvement if relaxation is suppressed.
\begin{figure}[t]
\vspace{-0.1in}
\hspace*{-0.2cm}
\includegraphics[width=9.2cm,height=11.5cm]{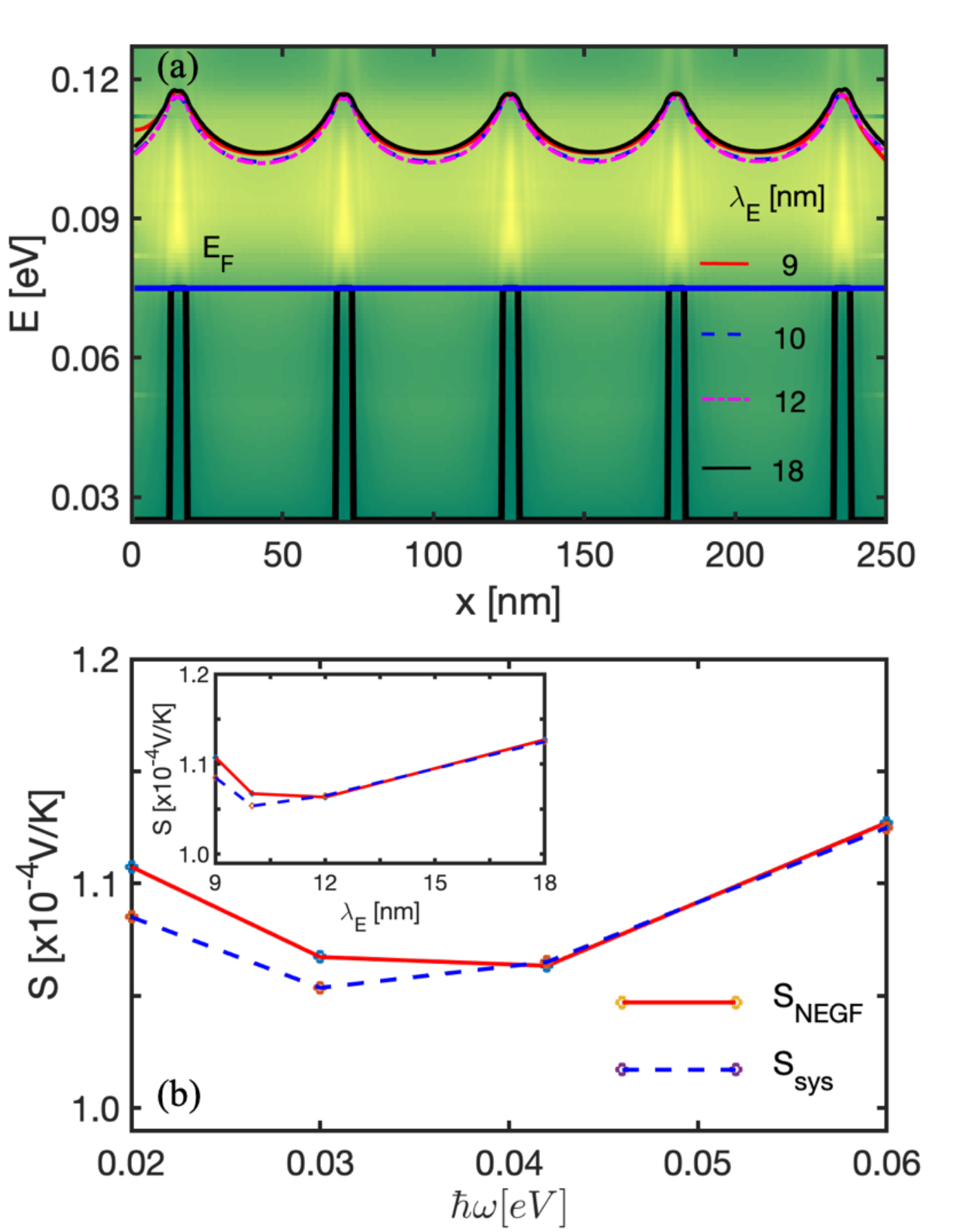}\\
\vspace*{0cm} \caption{\label{fig8} (Colour online)
(a) Average energy of the current flow $\langle E(x) \rangle$ along the channel with five barriers calculated from Eq.~(\ref{eq5}) using NEGF (solid, red line) and from the analytical result Eq.~(\ref{eqA3}) (dashed, blue line) for which $d = 50$nm and $\lambda_\text{E} = 17.5$nm. Here $E_\text{F} = V_\text{B} = 0.075$eV. The color map indicates the current flow $I(E,x)$, with yellow indicating high, and green low current density. (b) Seebeck coefficient vs grain size $d$ calculated from (i) NEGF (solid, red line), (ii) model with energy relaxation (dashed, blue line), and (iii) model without energy relaxation (dashed-dotted, magenta line).
}
\end{figure}

The relaxation length is again extracted graphically by fitting Eq.~(\ref{eqA5}) on the NEGF simulation result, which now yields $\lambda_\text{E} = 17.5$nm. Also, in the same manner as in the previous case, we find $\langle E \rangle_\text{B} = 0.117$eV and $\langle E \rangle_\text{W} = 0.096$eV, which yield $S_\text{B} = 1.4 \times 10^{-4}$V/K and $S_\text{W} = 0.7 \times 10^{-4}$V/K. In Fig.~7(a) we show the average energy of the current flow $\langle E(x) \rangle$ calculated from NEGF (solid, red line) and that from the model, i.e., Eqs.~(\ref{eqA1}) and (\ref{eqA3}) (dashed, blue line) in the case of five barriers. The rate of energy relaxation in this case is significantly smaller than in the previous case of Sec.~IIIA where the Fermi level was at $E_\text{F} = V_\text{B} - k_\text{B} T$ [compare with Fig.~3(a)]. Indeed, $\langle E \rangle_\text{B} - \langle E \rangle_\text{W} = 0.023$eV when $E_\text{F} = V_\text{B}$, while $\langle E \rangle_\text{B} - \langle E \rangle_\text{W} = 0.039$eV when $E_\text{F} = V_\text{B} - k_\text{B} T$, i.e., more than $40\%$ decrease. The reason for the suppressed relaxation in this case where $V_\text{B} = E_\text{F}$ compared to the previous one where $V_\text{B} > E_\text{F}$, is simply because electrons in the wells now tend to relax at the Fermi level (which is at the barrier level) and not below the barrier level (although the Seebeck coefficients are lower now due to the higher $E_\text{F}$. The variation of the Seebeck coefficient with $\hbar \omega$ is in this case smaller as well, since the current flow is closer to $E_\text{F}$, and out-scattering to filled lower energies away from $E_\text{F}$ is more difficult.

In Fig.~7(b) we show the Seebeck coefficient $S_{\text{NEGF}}$ calculated from NEGF (solid, red line) as a function of well size $d$, and the Seebeck coefficient $S_{\text{sys}}$ calculated from the model with energy relaxation Eq.~(\ref{eq16}) (dashed, blue line). The dashed-dotted magenta line shows the Seebeck coefficient $S_{\text{sys}}^{\text{no rel}}$ from the model without energy relaxation Eq.~(\ref{eq3}). The bulk values of the Seebeck coefficients $S_{\text{B}}$ and $S_{\text{W}}$ are also shown for reference (dashed-dotted, black likes). Again, as in the previous case of Fig.~3, in these calculations $d$ increases by removing barriers sequentially one at a time. We note the very good agreement of $S_{\text{sys}}$ with $S_{\text{NEGF}}$ to an accuracy of up to $\approx 5\%$ for large $d$. We also note that, as a consequence of the slower rate of energy relaxation, the Seebeck coefficient also decreases at a slower rate with increasing $d$ than in the previous case illustrated in Fig.~3(b).

\subsection{$S$ vs $\hbar \omega$ ($E_\text{F} = V_\text{B}$)}

We vary now the optical phonon energy $\hbar \omega$, while keeping the deformation potential fixed at $D_\text{0} = 0.0026$eV$^2$. We expect that the average energy of the current flow and the Seebeck coefficient exhibit the same behaviour as in the previous case. In Fig.~8(a) we show the average energy of the current flow $\langle E(x) \rangle$ calculated with NEGF for increasing values of $\lambda_\text{E}$. These values were extracted from fitting of Eq.~(\ref{eqA5}) to the NEGF result for each value of $\hbar \omega$ as we have done in Sec.~IIIC and, in addition, each value of $\langle E \rangle_\text{W}$ was determined as in Sec.~IIIC. We notice the small effect of $\hbar \omega$, which is also reflected in the Seebeck coefficient. This is shown in Fig.~8(b) where it is also seen that $S_\text{sys}$ agrees very well with $S_{\text{NEGF}}$ to an accuracy of $1-2\%$ for smaller values of $\hbar \omega$. We also notice that the Seebeck coefficient exhibits identical behaviour as that shown in Sec.~IIIC [see Fig.~5(b)] except the small variation with $\hbar \omega$.

\section{Comparison to Experiment}

To partially validate the Seebeck model with energy relaxation we constructed, we use the measured data in the experiment of Ref.~\cite{Zeng07,Bian07}, for the case of the Seebeck coefficients in SLs based on ErAs doped InGaAs wells and InGaAlAs barriers. The papers provide the estimated band offsets of the wells and barriers compared to the position of the Fermi level, as well as the measured Seebeck coefficients for the in-plane and cross-plane directions. Although we do not have access to other necessary parameters to compute electronic transport reliably in correlation with the experiment, at first order we can still approximate the Seebeck coefficient using Boltzmann Transport theory under the relaxation time approximation as:
\begin{figure}[t]
\vspace{-0.1in}
\hspace*{-0.3cm}
\includegraphics[width=9.4cm,height=6.2cm]{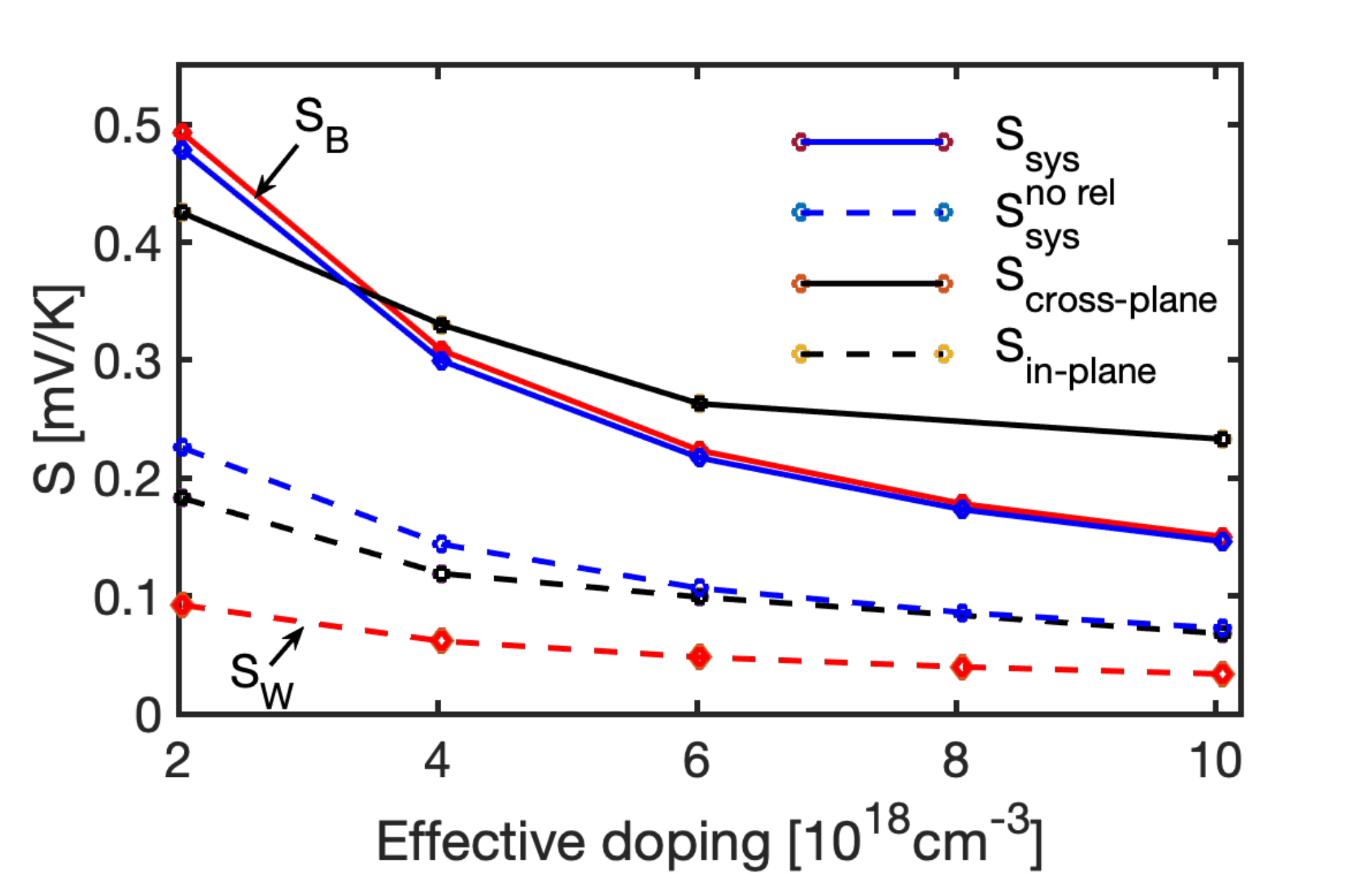}\\
\vspace*{0cm} \caption{\label{fig9} (Colour online)
Comparison between experimental and theoretical Seebeck coefficients for ErAs:InGaAs/InGaAlAs SLs with different doping concentrations from Ref.~\cite{Zeng07}. The black-solid and black-dashed lines show measured data for cross-plane and in-plane Seebeck coefficients in the SLs. The red-solid and red-dashed lines show the theoretically estimated Seebeck coefficients of the uniform well $S_{\text{W}}$ and barrier $S_{\text{B}}$, respectively. The blue-solid and blue-dahed lines show the theoretical calculations for the model that accounts for the relaxations physics, and the model that considers independent regions, respectively.
}
\end{figure}
\begin{equation}
S = \frac{q k_{\text{B}}}{\sigma} \int_{E_{\text{0}}}^{\infty} d E \left( -\frac{\partial f_{\text{0}}}{\partial E} \right) \Xi ( E ) \left( \frac{E - E_{\text{F}}}{k_\text{B} T} \right),
\label{eq18}%
\end{equation}
where
\begin{equation}
\sigma = q^2 \int_{E_{\text{0}}}^\infty d E \left( - \frac{\partial f_{\text{0}}}{\partial E} \right) \Xi ( E ),
\label{eq19}%
\end{equation}
with the transport distribution function $\Xi(E)$ defined as:
\begin{equation}
\Xi(E) = \tau (E) \upsilon (E)^2 g(E),
\label{eq20}%
\end{equation}
where $\upsilon (E)$ is the carriers' velocity, $g(E)$ is the density of states, and $\tau (E)$ is the relaxation time. Notice that $\upsilon(E)^2 \sim E$, while in 3D $g(E) \sim E^{1/2}$, and it is common to express the energy dependence of the relaxation times for acoustic phonons and ionized impurity scattering as $\sim E^{-1/2}$ and $\sim E^{3/2}$, respectively. In the experiment a series of doping values in the $10^{18}$ - $10^{19}$cm$^{-3}$ range were used, thus, we employ a mixed scattering relaxation time exponent in BTE as $r = 1/2$ \cite{Mao17}. Thus, the Seebeck coefficients for the well $S_{\text{W}}$ and the barrier $S_{\text{B}}$ can be approximated using the band edges provided in the experimental paper as:
\begin{equation}
S = \frac{k_\text{B}}{q} \frac{\int_{E_0}^\infty E^2 \left( - \frac{\partial f_0}{\partial E} \right) \left( \frac{E - E_{\text{F}}}{k_{\text{B}} T} \right) d E}{\int_{E_0}^\infty E^2 \left( -\frac{\partial f_0}{\partial E} \right) d E},
\label{eq21}%
\end{equation}
where now at first order it does not depend on material parameters. In this structure the barrier thickness is $b = 10$nm, while the well length is $d = 20$nm, and the only parameter needed is the relaxation length, which is taken to be $\lambda_{\text{E}} = 30$nm, to reflect the higher mobility of InGaAs compared to Si.

Figure 9 shows the measured data for the in-plane and cross plane Seebeck coefficients by the black-dashed and black solid lines, respectively versus carrier density. The red-dashed and red-solid lines show the uniform channel calculated $S_\text{W}$ and $S_\text{B}$ (upper and lower limits of our calculations). The blue-dashed line shows the calculated Seebeck coefficients in the case where each region in considered to be independent (no relaxation physics considered), whereas the blue-solid line when the relaxation physics is considered by the model developed. Despite the large uncertainties of this evaluation, the model (blue-solid line) is in the range of the measured cross-plane data (black-solid line). We find that slightly better fit can be obtained by adjusting the scattering time exponents, however, we do not attempt better fit as any exponent we use will be purely speculative. On the other hand, when the barrier and well are considered to be independent, the independent region model resides lower and coincides with the in-plane measured Seebeck data (although this could just be accidental). Despite the large uncertainties, this analysis shows the validity of the developed model in describing the Seebeck coefficient of SLs and nanocomposite systems.

\section{Summary and Conclusions}

In summary, we presented a simple analytical model for the Seebeck coefficient $S$ of superlattice materials (or nanocomposites to first order approcimation) in the presence of energy relaxation due to electron-optical phonon scattering. This model casts the complex and crucial physics of semi-relaxation and its role in Seebeck enhancement in terms of only three material parameters: the bulk Seebeck coefficients of the constituent materials and the energy relaxation length $\lambda_\text{E}$, which is related to the energy relaxation time $\tau_\text{E}$, of the more conductive ``well'' material. Thus, it is our hope that the model can help guide future nano-engineering efforts aimed at Seebeck coefficient and power factor enhancement.

To validate this model, numerical simulations were performed using the fully quantum mechanical nonequilibrium Green's function method and very good agreement was found. We also compared the model with experimental values for ErAs:InGaAs/InGaAlAs superlattice, and good agreement was found. The variation of the calculated Seebeck coefficient with increasing well size $d$ and Fermi level $E_\text{F}$, as well as with increasing e-ph coupling strength $D_\text{0}$, and with increasing optical phonon energy $\hbar \omega$, the physical parameters that determine $\lambda_\text{E}$, were also studied. We also provided an expression for the average energy of the current flow $\langle E(x) \rangle$, which agrees very well with the NEGF result and captures accurately the behavior of $\langle E(x) \rangle$ in the well regions. We expect that these results will be helpful and useful to experimentalists in their determination of the Seebeck coefficients of nanocomposite structures and superlattices. Note that the paper deals exclusively with the Seebeck coefficient because it can be trivially mapped to the average energy of the current flow. Similar considerations for the electrical conductance in the semi-energy relaxing regions between the barriers and the wells are more complicated, as there is no direct map to the average energy of the current flow.

Finally, we note that in all simulations of this work and in the construction of the model, we considered periodic superlattice structures. However, we argue that the model is at first order applicable to nanocomposite/nanocrystalline materials as well. Nanocomposites are described by a 3D aperiodic geometry, and strictly speaking the complexity of the transport paths is such that would not allow us to map the 3D onto 1D paths beyond a first order estimation. Superlattice geometries thought, can be considered as a limiting case for a
nanocomposite system, becoming more accurate as the variance of barrier spacing and size decreases and the structure becomes more uniform in shape and distribution. Indeed, in previous works of ours \cite{Thes15JAP,ThessJEM16} we pointed out that in the presence of statistical variability in the sizes of the domains, the overall Seebeck coefficient is rather robust. Thus, we still believe that the model developed provides a first order estimate to the Seebeck coefficient of 3D nanocomposites and nanocrystalline structures as well. \\

{\bf Acknowledgments} This work has received funding from the European Research Council (ERC) under the European Union's Horizon 2020 Research and Innovation Programme (Grant Agreement No. 678763). We also thank Samuel Foster and Dhritiman Chakraborty for helpful discussions.\\

\appendix

\section{Average energy of the current flow $\langle E(x) \rangle$ for SL structures}

We consider a 1D SL structure of length $L$, as shown in Fig.~1(a). We make the approximation of constant $\langle E(x) \rangle$ over each barrier, i.e.,
\begin{equation}
\langle E(x) \rangle = \sum_{\ell=0}^{n-1} \langle E \rangle_{\text{B}} \Theta (x - x_0 - \ell t) \Theta (x_0 + b + \ell t -x) ,
\label{eqA1}%
\end{equation}
where $\langle E \rangle_{\text{B}}$ is the average energy on the barrier, $\Theta(...)$ is the Heaviside function, $n$ is the total number of barriers, and $t = b + d$. However, within a well, $\langle E(x) \rangle$ decays exponentially \cite{BMoyz98}. For simplicity, we consider one well with length $d$, surrounded by two barriers at $x=0$ and $x=d$. In addition to the exponential decay, $\langle E(x) \rangle$ should satisfy two boundary conditions, i.e., $\langle E(x) \rangle = \langle E \rangle_{\text{B}}$ at $x=0$ and $x=d$. The expression for $\langle E(x) \rangle$ should also satisfy the equilibrium condition for large $d$, i.e., $\langle E(x) \rangle = \langle E \rangle_{\text{W}}$ at $x = d/2$ as $d \rightarrow \infty$, where $\langle E \rangle_{\text{W}}$ is the average energy in the well region under equilibrium. This means that as the well size becomes large (i.e., $d \gg \lambda_{\text{E}}$), the energy of the charge carriers is fully relaxed within the well and $\langle E(x) \rangle$ reaches the bulk limit. The differential equation that yields solutions which satisfy these conditions is of second order and given as
\begin{equation}
\frac{d^2 \langle E \rangle}{d x^2} - \frac{\langle E \rangle}{\lambda_{\text{E}}^2} = \frac{1}{\lambda_{\text{E}}^2} \left( \langle E \rangle_{\text{B}} - \langle E \rangle_{\text{W}} \right) e^{-d / \lambda_{\text{E}}} - \frac{\langle E \rangle_{\text{W}}}{\lambda_{E}^2} .
\label{eqA2}%
\end{equation}
The solution of Eq.~(\ref{eqA2}) proceeds in a straightforward manner and is given as
\begin{eqnarray}
\nonumber \hspace{-0cm} \langle E(x) \rangle = \sum_{\ell=0}^{n-2} ( \langle E \rangle_{\text{B}} - \langle E \rangle_{\text{W}} )  [ e^{-(x - x_0 - b - \ell t ) / \lambda_{\text{E}}}  
\\* &&\hspace*{-2.4in} + e^{(x - x_0 - (\ell+1) t ) / \lambda_{\text{E}}} - e^{-d / \lambda_{\text{E}}}  ] + \langle E \rangle_{\text{W}} .
\label{eqA3}%
\end{eqnarray}
In Eq.~(\ref{eqA3}) we notice that as $d$ becomes large, $\langle E(x) \rangle$ gradually relaxes, and $\langle E(d/2) \rangle \rightarrow \langle E \rangle_\text{W}$ as $d \rightarrow \infty$. 

In the wells at the channel ends, i.e., close to the contacts, $\langle E(x) \rangle$ satisfies different boundary conditions. For simplicity, we consider the well at the right channel end and we make the transformation $x \rightarrow x - x_0 - (n-1) t + b$. Then, $\langle E(x) \rangle$ should only satisfy the boundary conditions $\langle E(x) \rangle = \langle E \rangle_\text{B}$ at $x=0$ and, in addition, the equilibrium condition $\langle E(x) \rangle = \langle E \rangle_\text{W}$ at $x = d^{\prime} / 2$ as $d^{\prime} \rightarrow \infty$. The differential equation that yields solutions which satisfy these conditions is of first order and given as
\begin{equation}
\frac{d \langle E \rangle}{d x} + \frac{\langle E \rangle}{\lambda_\text{E}} = \frac{\langle E \rangle_\text{W}}{\lambda_\text{E}}  .
\label{eqA4}%
\end{equation}
The presence of the term on the right hand side of Eq.~(\ref{eqA4}) guarantees that the solution satisfies the equilibrium condition. The solution of Eq.~(\ref{eqA4}) is given as
\begin{equation}
\langle E(x) \rangle = \left( \langle E \rangle_\text{B} - \langle E \rangle_\text{W} \right) e^{- x / \lambda_\text{E}} + \langle E \rangle_\text{W}  .
\label{eqA5}%
\end{equation}
The solution Eq.~(\ref{eqA5}) is relevant to the well in the right channel end. For the well in the left channel end, we just make the replacement $x \rightarrow d^{\prime} - x$ in Eq.~(\ref{eqA5}).




\begin{thebibliography}{9}

\bibitem{GDing16} G.~Ding, G.~Y.~Gao, Z.~Huang, W.~Zhang and K.~Yao, Nanotechnology \textbf{27}, 375703 (2016).

\bibitem{HHuang16} H.~Huang, Y.~Cui, Q.~Li, C.~Dun, W.~Zhou, W.~Huang, L.~Chen, C.~A.~Hewitt, and D.~L.~Carroll, Nano Energy \textbf{26}, 172 (2016). 

\bibitem{WHuang14} W.~Huang, X.~Luo, C.~K.~Gan, S.~Y.~Quek, and G.~Liang, Phys.~Chem.~Chem.~Phys. \textbf{16}, 10866 (2014).

\bibitem{Beek15nmat} M.~Beekman, D.~T.~Morelli, and G.~S.~Nolas, Nat.~Mater. \textbf{14}, 1182 (2015).

 \bibitem{CFu15ncom} C.~Fu, S.~Bai, Y.~Liu, Y.~Tang, L.~Chen, X.~Zhao, and T.~Zhu, Nat.~Commun. \textbf{6}, 8144 (2015). 
 
 \bibitem{Biswas12} L.-D.~Zhao, S.-H.~Lo, Y.~Zhang, H.~Sun, G.~Tan, C.~Uher, C.~Wolverton, V.~P.~Dravid, and M.~G.~Kanatzidis, Nature (London) \textbf{508}, 373 (201). 

\bibitem{Beretta19} D.~Beretta, N.~Neophytou, J.~M.~Hodges, M.~G.~Kanatzidis,
D.~Narducci, M.~M.- Gonzalez, M.~Beekman, B.~Balke, G.~Cerrettii, W.~Tremel \textit{et al.,} "Thermoelectrics: From history, a window to the future," Mater.~Sci.~Eng.~R Rep. (to be published).
 
\bibitem{Mizuno15} H.~Mizuno, S.~Mossa, and J.-L.~Barrat, Sci.Rep. \textbf{5}, 14116 (2015).
 
\bibitem{Thes15JAP} M.~Thesberg, M.~Pourfath, H.~Kosina, and N.~Neophytou, J.~Appl.~Phys.~\textbf{118}, 224301 (2015). 

\bibitem{Thes16JAP} M.~Thesberg, H.~Kosina, and N.~Neophytou, J.~Appl.~Phys. \textbf{120}, 234302 (2016).

\bibitem{MZhou17} Y.~M.~Zhou and L.-D.~Zhao, Adv.~Mater. \textbf{29}, 1702676 (2017).

\bibitem{Priy18} P.~Priyadarshi, A.~Sharma, S.~Mukherjee, and B.~Muralidharan, J.~Phys.~D: Appl.~Phys. \textbf{51}, 185301 (2018).

\bibitem{Venk96} R.~Venkatasubramanian, T.~Colpitts, E.~Watko, and J.~Hutchby, in Proceedings IEEE 15th International Conference on Thermoelectrics (IEEE Service Centre, 1996).

\bibitem{Ishida09} A.~Ishida, T.~ Yamada,  D.~Cao, Y.~Inoue, M.~Veis, and T.~Kita, J.~Appl.~Phys. \textbf{106}, 023718 (2009).

\bibitem{Zeng07} G.~Zeng, J.~M.~O.~Zide, W.~Kim, J.~E.~Bowers, A.~C.~Gossard, Z.~Bian, Y.~Zhang, A.~Shakouri, S.~L.~Singer, and A.~Majumdar, J.~Appl.~Phys. \textbf{101}, 034502 (2007).

\bibitem{Bian07} Z.~Bian, M.~Zebarjadi, R.~Singh, Y.~Ezzahri, A.~Shakouri, G.~Zeng, J.-H.~Bahk, J.~E.~Bowers, J.~M.~O.~Zide, and A.~C.~Gossard, Phys.~Rev.~B \textbf{76}, 205311 (2007).

\bibitem{Neo13} N.~Neophytou, X.~Zianni, H.~Kosina, S.~Frabboni, B.~Lorenzi, and D.~Narducci, Nanotechnology \textbf{24}, 205402 (2013).
 
\bibitem{Vargiam19} V.~Vargiamidis and N.~Neophytou, Phys.~Rev.~B \textbf{99}, 045405 (2019).

\bibitem{NeoJAP13} N.~Neophytou and H.~Kosina, J.~Appl.~Phys. \textbf{114}, 044315 (2013).

\bibitem{Popescu09} A.~Popescu, L.~Woods, J.~Martin, and G.~Nolas, Phys.~Rev.~B \textbf{79}, 205302 (2009). 

\bibitem{CBera10} C.~Bera, M.~Soulier, C.~Navone, G.~Roux, J.~Simon, S.~Volt, and N.~Mingo, J.~Appl.~Phys. \textbf{108}, 124306 (2010).

\bibitem{MBart01} M.~Bartkowiak, G.~D.~Mahan, and M.~T.~Terry, in \textit{Semiconductors and Semimetals} (Elsevier, 2001), Vol.~70, pp. 245-271.

\bibitem{Koswatta07} S.~O.~Koswatta, S.~Hasan, M.~S.~Lundstrom, M.~P.~Anantram, and D.~E.~Nikonov, IEEE Trans.~Electron Devices \textbf{54}, 2339 (2007).

\bibitem{Anantram08} M.~P.~Anantram, M.~Lundstrom, and D.~Nikonov, Proc.~IEEE \textbf{96}, 1511 (2008).

\bibitem{Kim11} R.~Kim and M.~Lundstrom, J.~Appl.~Phys.~\textbf{110}, 034511 (2011).

\bibitem{Nard15} X.~Zianni and D.~Narducci, J.~Appl.~Phys. \textbf{117}, 035102 (2015).

 \bibitem{Lundstrom} M.~Lundstrom, \textit{Fundamentals of Carrier Transport}, Cambridge, UK: Cambridge Univ. Press, 2000.
 
\bibitem{Selberherr} B.~Gonzalez, V.~Palankovski, H.~Kosina, A.~Hernandez, and S.~Selberherr, Solid-State Electron. \textbf{43}, 1791 (1999).

\bibitem{Kim12} R.~Kim and M.~Lundstrom, J.~Appl.~Phys.~\textbf{111}, 024508 (2012).

\bibitem{BMoyz98} B.~Moyzhes and V.~Nemchinsky, Appl.~Phys.~Lett. \textbf{73}, 1895 (1998).

 \bibitem{Neop14} N.~Neophytou, X.~Zianni, H.~Kosina, S.~Frabboni, B.~Lorenzi, and D.~Narducci, J.~Elec.~Mater. \textbf{43}, 1896 (2014).
 
\bibitem{ThesPRB17} M.~Thesberg, H.~Kosina, and N.~Neophytou, Phys.~Rev.~B \textbf{95}, 125206 (2017).

\bibitem{Seebeck} In experimental settings, one extracts the Seebeck coefficient from the open circuit voltage upon the application of a thermal gradient along the channel, as $S = \Delta V / \Delta T$, which equivalently can also be computed by $S = I_{(\Delta V = 0)} / G \Delta T$. In \cite{Kim11}, it was validated that the two methods of extracting the Seebeck coefficient are equivalent, which makes it easier in time consuming simulations (as the ones we undertake) to only run the $\Delta V \neq 0$ case and still be able to extract the Seebeck coefficient by integrating the energy of the current flow over the length of the channel.

\bibitem{Mao17} J.~Mao, J.~Shuai, S.~Song, Y.~Wu, R.~Dally, J.~Zhou, Z.~Liu, J.~Sun, Q.~Zhang, C.~dela Cruz, S.~Wilson, Y.~Pei, D.~J.~Singh, G.~Chen, C.-W.~Chu, and Z.~Ren, PNAS \textbf{114}, 40, 10548 (2017).

\bibitem{ThessJEM16} M.~Thesberg, M.~Pourfath, N.~Neophytou and H.~Kosina, J.~Electron.~Mater. \textbf{45}, 1584 (2016).

\end{thebibliography}
\end{document}